\documentclass[letterpaper,12pt]{article}
\usepackage{osajnl2}
\begin{document}

\title{Time-resolved quantitative multiphase interferometric imaging of a highly focused ultrasound pulse}

\author{Fabien Souris,$^{1}$  Jules Grucker,$^{1}$ Jacques Dupont-Roc,$^{1}$  Philippe Jacquier,$^{1}$
Arnaud Arvengas,$^{2}$ and Fr\'ed\'eric Caupin,$^2$}
\address{$^1$Laboratoire Kastler Brossel, ENS, UPMC-Paris 6, CNRS \\ 24 rue Lhomond, 75005 Paris, France}
\address{$^2$Laboratoire de Physique Statistique, ENS, UPMC-Paris 6, Universit\'e Paris Diderot, CNRS \\ 24 rue Lhomond, 75005 Paris, France}
\address{$^*$Corresponding author: fabien.souris@lkb.ens.fr}

\begin{abstract}
Interferometric imaging is a well established method to image phase objects by mixing the
image wavefront with a reference one on a CCD camera. It has also been applied to fast transient phenomena, mostly through
the analysis of single interferograms. It is shown that for repetitive phenomena multiphase acquisition
brings significant advantages. A $1$~MHz focused sound field emitted by a hemispherical piezotransducer in water is imaged as
an example. Quantitative image analysis provides high resolution sound field profiles. Pressure
at focus determined by this method agrees with measurements from a fiber-optic probe hydrophone. This confirms that 
multiphase interferometric imaging can indeed provide quantitative measurements.
\end{abstract}


\maketitle 
\section{Introduction}\label{intro} %
Interferometric imaging is a common method to investigate  phase objects (see for instance \cite{creath1988,
schwider1990, vlad1994} and ref. therein). Sound waves are examples of such objects. Compared to other imaging
methods such as Schlieren imaging (see for instance \cite{willard1949}), or stress induced birefringence
imaging\cite{ying1990}, interferometric imaging has the advantage of providing directly quantitative
measurements. Refractive index maps can be computed from those measurements, and in the case of a sound wave, acoustic
pressure can be determined. There are two methods to reconstruct phase maps from recorded interferograms.
Fringe analysis \cite{takeda1982,schwider1990} has the advantage of requiring only one image (or two if
fringe shift is used), but implies a reduced spatial resolution and phase ambiguities. In multiphase
(or phase shifting) interferometry, the map of the optical phase is extracted from several images taken
while changing the phase of the reference beam. Applying this method to time dependent phenomena requires
special adaptations to circumvent the slow acquisition rate of common cameras. When the phenomenon is a
steady sinusoidal oscillation with sufficient amplitude, information can be extracted from time averaged
images\cite{ezure2004}. This is the case for the popular TV holography method\cite{lokberg1994}. For fast
transient phenomena, a time sampling is to be made by using pulsed light sources to get images of the object
at particular times. These quasi-instantaneous images registered by the CCD camera are read on much longer
times. Recording the complete evolution is possible for repetitive transient phenomena by varying the
delay between the laser pulse and the triggered phenomenon. Depending on the time scale of the phenomenon
under study, pulsed light sources  such as pulsed LED\cite{gutierrez1999, conway2007}, Q-switched
lasers\cite{kwon1987, harris1979}, or even femtosecond laser\cite{Abraham2000, Gahagan2002, Temnov2004,
Funk2005} have been used.

Regarding multiphase interferometric imaging for fast transient phenomena, the litterature provides very
few examples. Fringe analysis with tilted reference wavefronts seems to be the widely used method, with the
drawbacks mentionned previously. It is thus of interest to investigate whether pulsed multiphase
interferometry can provide phase maps, or phase shift maps, for fast transient phase objects. In this
article, it is shown that this method can indeed be used in such cases, and provides accurate images with
little sensitivity to optical defects. The price to be paid is a longer acquisition process.  As an example,
a 1~MHz ultrasound pulse  produced in water by an hemispherical transducer is imaged. In section~2, the
experimental arrangement is described as well as the procedure to record several images with different
reference phases, at various time delays. In section~3, the corresponding data processing is detailed. The phase maps are
extracted, and unwrapped if phase shifts exceed $\pm \pi$. Then an inverse Abel transformation is applied to
recover the local index of refraction, and hence the sound pressure. The result is shown in section~5 and
discussed in section~6.

\section{Experimental setup and procedure}\label{sec:2}
The experimental setup is sketched in Figure 1. It is based on a Jamin interferometer\cite{born} with a
frequency doubled pulsed Nd:YAG laser as a light source. The Jamin interferometer was chosen because of its
simplicity and intrinsic stability. When separated, the two interfering beams are close to each other and
undergo similar disturbances from air turbulence and encountered windows\cite{bt}. In the sample region, the
two beams are 6~mm apart and are 6~mm high. The lens ${\rm L_1}$ images the entrance rectangular diaphragm ${\rm A}$ on the sample,
while ${\rm L_2}$ makes an image of the sample onto the CCD camera.
The laser (Big Sky Laser CFR200) delivers 120~mJ, 8~ns long pulses with a
repetition rate of 20~Hz, from which only about 1~$\mu$J pulses are derived by 4 vitrous reflections, and a
$d=2$ gray filter. The optical wavelength used is $\lambda_o= 532$~nm. The laser is transversally highly
multimode with a flat top profile. Its effective coherence length, as measured with the interferometer by the
fringe contrast versus the optical path difference, was found to be 4~mm at half contrast. Hence the optical
path difference between the two beams was kept well below this value to get a good contrast.
The imaging camera (Andor LUCA~R) is triggered synchronously with the pulsed laser. Used with a $2\times 2$
binning, it provides $280\times 502$ pixel images of the $6\times 10.5$ mm$^2$ observed region. Image
resolution  is discussed at the end of this section and is of about 20~$\mu$m.

For scanning the interferometer, two 4~mm thick parallel plates ${\rm S_1}$ are introduced in the reference beam,
with symmetrical tilts to avoid beam displacement (see Fig. 1). They are allowed to rotate around an
horizontal axis so that their tilt angle  can be precisely controlled around 5$^\circ$. This is achieved by
moving a separator along the vertical axis. When changing its heigth $h$ by $\delta h = 1$~mm, the tilt
angle is changed by $0.05^\circ$, and the optical path by $0.14$~$\mu$m. Because of their thickness, the
plates ${\rm S_1}$ also introduce a difference of about 4~mm in the optical length of the two interferometer paths.
In order to keep the optical path difference smaller than the laser coherence length, a second pair of
similar plates ${\rm S_2}$ is introduced in the signal beam. Those are kept fixed. The phase shift between the two
beams introduced by these devices is approximated as:
\begin{equation}\label{psi}
  \psi = \psi_0 + 2\pi {h\over a} \left(1 + s {h\over a} \right)
\end{equation}
where $a \simeq 3.7$~mm and $s \simeq 0.04$ as determined by simple geometrical optics. In fact the two
parameters $a$ and $s$ depend somewhat on the incidence angle of the beams on the plates, which cannot be
measured easily. So they are left as fitting parameters as explained later. 

The phase object to be imaged is an ultrasound wave pulse (central frequency 1.06 MHz, duration 5~$\mu$s)
emitted in water along the vertical axis by a  piezoelectric hemisphere (inner diameter 12~mm, outer diameter 16~mm,
provided by Channel Industries Inc.). Water used was filtered by reverse osmosis (Millipore Direct-Q ultrapure 
water system).  The thickness vibration mode of the transducer is driven by an RF amplifier (RF in Fig. 1)
fed with a function generator (Tektronix AFG3022) triggered from the laser command electronics
with an adjustable delay. The same generator also triggers image acquisition by the camera. The sound
velocity in water is 1480~m/s \cite{fine1973} at $T=293$~K. The sound wavelength is $\lambda_s \simeq 1.4$~mm and the
sound pulse length about $L_s=7$~mm. The sound pulse time of flight from the transducer inner surface to
its center is about 4~$\mu$s. In the experiment trigger is given by the laser flash firing. The
laser pulse takes place 170~$\mu$s later with a negligible jitter. The sound pulse is sent at time $t$ varied
from 157 to 169~$\mu$s by steps of 0.01~$\mu$s ($N_t=1200$ time steps). For each time step, images are taken
for $N_p = 25$ different values of $h$, labeled $h(1) ... h(p)... h(N_p)$. The corresponding values of $\psi$, labeled $\psi (1) ... \psi (p) ... \psi (N_p)$, span an interval set to  $\psi (N_p) - \psi (1)\simeq 3\pi$. Raw data are a set of $N_p\times N_t$ images labelled with their time delay and phase step. Their
recording takes about $30$ minutes. For each phase step, a reference image is first made by averaging 
$N_{av}=20$ images without exciting the transducer. These images  of an undisturbed medium are later used
as reference images for subtraction of a background phase field. The order in which $h$ and $t$ are varied
is important to minimize interferometer drift effects. The sound pulse insertion time $t$ is varied
first at a fixed phase step along a complete time series, before changing $h$ to the next phase step.

\section{Phase map computation}\label{sec:3}
During data processing, pixels are treated independently. Hence there is no spatial
resolution loss due to image processing. Let $I(t,p)$ be the intensity recorded for a given pixel for the
time index $t$ and interferometer phase index $p$. It is expected that \cite{creath1988}: %
\begin{equation}\label{interf}
I(t,p)=I_0(t) \left[1+C(t) \cos(\phi(t) - \psi(p))\right]
\end{equation}
where $I_0(t)$ is the average intensity, $C(t)$ the fringe contrast, $\phi(t)$ the optical phase to be
measured for this particular pixel and $\psi(p)$ is the interferometer phase given by equation (\ref{psi}) for
each value of $h$. To extract  $\phi(t)$, one needs to fit the $N_p$ values $I(t,p)$ with equation (\ref{interf})
with five unknown parameters $I_0(t)$, $C(t)$, $\phi(t)-\psi_0$, $a$, $s$. Because of the nonlinear term
$(h/a)^2$ in equation (\ref{psi}), this involves a time consuming nonlinear fitting procedure. Actually the
parameters $\psi_0$, $a$ and $s$ do not depend much on time and position. Thus they are only computed for pixels
of the reference images for which $\phi(t)$ is zero. Actually the variance of $a$ is a few percent. Hence, only
a mean value over space of those three parameters is computed and used in the next steps of the
process.
In the later images  only three parameters are to be determined for each pixel, namely $I_0(t)$,
$C(t)\cos(\phi(t))$ and $C(t)\sin(\phi(t))$. This can be done with a fast linear
method \cite{morgan1982, creath1988} and takes about 1~s for the entire image of one time step.

The value $\phi(t)$ determined in this way is within the limits $-\pi$, $\pi$ and undergoes possibly $2\pi$
phase jumps. The first time step is chosen so that the phase shift amplitude introduced by the sound field is less
than $\pi$. The phase unwrapping is realized in time for each pixel by removing phase jumps $|\phi(t+1)-\phi(t)|$
larger than $\pi$. This is possible only if time steps verify the sampling condition, i.e.
$|\phi(t+1)-\phi(t)|< \pi - 2\delta\phi$, where $\delta\phi$ is the phase noise. This condition
actually sets the value of the time step and consequently the number of steps in the experiment. Usually phase
unwrapping is made spatially for each time on the 2D-maps, and then made continuous in time. Although many
algorithms have been published on this topic\cite{chyou2004}, it is not a trivial problem. Unwrapping the
phase in time for each pixel separately is much easier and faster. The continuity in space for the phase shift map 
$\phi(t,x,z)-\phi(0,x,z)$ is obtained automatically, because it starts from 0 everywhere 
and has no later phase jumps.

\section{From phase map to radial sound field}\label{sec:4}
The phase $\phi$ is assumed to be related to the refractive index by a simple integration along the $y$-axis
parallel to the beam direction in the sample (see Figure 2). More precisely, for each pixel $(x,z)$, one takes into
account the difference $\delta n(x,z,y)$ between the perturbed index $n(x,z,y)$ seen by the signal beam at
point $(x,y,z)$ and the refractive index $n_0$ of the unperturbed fluid seen by the reference beam on its
parallel path. In our case the sound field is rotationnally invariant around the hemisphere axis $z$ so that
$\delta n$ is a function of $z$ and $r=\sqrt{x^2 + y^2}$ only. Then the phase map results from an integration of
the refractive index variations over the path of the beam from the entrance window ($y = -l/2$), to the exit
window ($y = l/2$):
\begin{equation}\label{phase}
  \phi(x,z) = {2 \pi\over \lambda}\int_{-l/2}^{l/2} {\rm d}y \ \delta n(\sqrt{x^2 + y^2},z).
\end{equation}
Note however that $\delta n$ is identically $0$ after some finite distance from the center (dashed circle on Figure 2)
because there is still no sound there. Hence equation (\ref{phase}) can be rewritten as:
\begin{equation}\label{phase_abel}
  \phi(x,z) = {2 \pi\over \lambda}\int_{-\infty}^{\infty} {\rm d}y \ \delta n(\sqrt{x^2 + y^2},z).
\end{equation}
which is the Abel transform of the $\delta n(r,z)$ map.
Conversely radial refractive index profiles can be retrieved from phase maps
via an inverse Abel tranform. To make that part easier, the camera lines and the signal light beam direction
$Oy$ have been carefully aligned parallel to the hemisphere basis. The abscissa $x_0$ of the hemisphere axis
is determined as the symmetry axis of the projected sound field. Then the  Abel inversion of the phase is perfomed
for each line $z$ on $\phi(z,x-x_0)$ using the algorithm proposed by \cite{deutsch1983}. The data are fitted with
splines over successive sets of 20 pixels.

Abel inversion procedure is justified provided that light rays do not suffer any significant deviation during their
propagation through the studied sample. More precisely the Raman-Nath condition should be fulfilled, i.e. deviation of a
light beam from a straight line over the sample diameter $D$ due to the refractive index gradient should be less
than the optical resolution $\delta x$. Let $\lambda_s$ be the sound wavelength, $\lambda_o$ the
laser light wavelength, $\phi_m$ the maximum phase accumulated across the sample. Then the condition amounts
to:
\begin{equation}\label{RamanNath}
  \phi_m < {\delta x \lambda_s \over D \lambda_o}.
\end{equation}
With the values $\lambda_o= 0.5$~$\mu$m, $\lambda_s=1.4$~mm, $\delta x \simeq \lambda_s / 10 $, condition
(\ref{RamanNath}) puts a rather large limit on $\phi_m$, about 50. A second condition is that all rays
collected by the imaging optics and originating from a given point of the object undergo the same $\phi$
retardation. Maximum transverse  extension of such a ray bundle at the exit of the sample, i.e. at $D/2$,
should be less than the characteristic length over which $\phi$ varies, typically $\lambda_s$. If $\theta$
is the aperture angle of the optics, this is true if $\theta D/2 <\lambda_s$. In our case, $D\simeq 16$~mm.
This leads to the condition $\theta < 0.2$. The numerical aperture of the imaging lens is about $\theta = 0.03$ and fulfills the
condition. In other respects, this numerical aperture should be large enough to ensure the desired optical
resolution $\delta x$. The current aperture provides a resolution $\delta x=1.2 \lambda_o /\theta \simeq 20\
\mu$m.

The acoustic pressure field is proportional to $\delta n$ and is computed from the relation:
\begin{equation}\label{pression}
  \delta P(x,z) = (n(x,z,y)-n_0)/ (\partial n/\partial P)
\end{equation}
where the water piezo-optic constant is $(\partial n/\partial P) \simeq 1.4 \times
10^{-4}$MPa$^{-1}$\cite{staudenraus1993, parsons2006}.

\section{An example of ultrasound pulse image}\label{sec:5}
In Figures 3 and 4, three stages are  shown from interference images to pressure map when the sound pressure
is at its maximum at the focus. On image 3.a, severe optical defects are clearly visible : dust particles,
diffraction by the hemisphere rim. These defects are nearly completely washed out in the phase map 3.b. On
the contrary, they could be seen obviously on the contrast map. This appears to be the main advantage of the
multiphase method, the fitted phase being only weakly sensitive to intensity fluctuations. The signal to
noise is good enough to yield an acceptable inverse  Abel transform, from which the pressure is deduced
using equation \ref{pression}. Its radial profile is shown in Figure 4.a, as well as its time variation at
the focus when the sound pulse is going through (Figure 4.b). While the tranducer is excited at its
resonance frequency by a constant amplitude sinusoidal burst of 5 periods, its response increases in time as
expected from a driven damped harmonic oscillator. The oscillation is actively damped during the last
oscillation. This explains the time profile of the sound pulse (Figure 4.b).

This imaging method thus provides a way to investigate in detail the focusing properties of this type of
transducer.

\section{Discussion}\label{sec:6}
Discussions of errors in multiphase interferometry may be found in \cite{morgan1982, schwider1983,
creath1988, schwider1990}. Although written for static interferometric imaging, they apply equally well to
the pulsed case. They show that the multiphase technique brings two benefits.  The first one is to avoid
systematic errors coming from imperfectly scanned phase $\psi_p$ when predetermined schemes such as $(-\pi
/2, 0, \pi /2)$, $(-\pi /2, 0, \pi /2, \pi)$ or $(-\pi, -\pi /2, 0, \pi /2, \pi)$ are used \cite{hariharan1987}. Here the phase
is not scanned with an {\em a priori} scheme. Its value is extracted from the series of unperturbed images at
$t=0$. In equation (\ref{psi}) the slope $2\pi /a$ of the actual phase as a function of the scanning
parameter $h$ is left free, and a non linearity $s$ is permitted by the fitting process. Hence the only 
remaining errors may come from drifts in time of the interferometer. Since in the experiment, changing the 
delay is ten time faster than changing the phase, images are taken for a complete time series at a fixed
reference phase. This takes about 60~s. Then the reference phase is changed to the next value. In this way
a possible drift in time of the interferometer geometry is in some way taken into account by the fitting
process of $a$ and $s$ in formula (\ref{psi}). This scanning procedure, i.e. \lq time first, phase second\rq ,
is thus to be preferred.

 The second benefit from multiphase technique results simply from the larger number of data (here 25 images
compare to a minimum of 3 to extract the phase, and compared to 1 for fringe analysis). This statistical
improvement is thus $\sqrt{25/3}$ and $\sqrt{25}$ respectively. More quantitatively, let $\delta I$ be the pixel
noise for one image (including laser fluctuations and detection noise):
\begin{equation}\label{dphi}
  \delta \phi = { {\delta I/I_0}\over \sqrt{ C  N_p} }.
\end{equation}
Here $\delta I/I_0\simeq 0.1$, $C \simeq 1$,  $N_p=25$. This gives  $\delta \phi \simeq 0.02$,
or $\lambda/300$ in a more commonly used unit.

Of course, the imaging technique has two well-known advantages compared to in situ point by point pressure
measurement with a pressure probe. It provides parallel measurements at a large number of locations and
times. This quality is reflected obviously by the large size of the generated data (GBytes). It also provides
this information in a non invasive and non perturbative way.

Nevertheless one may question the accuracy of the pressure field determined in this way. A quantitative
analysis of the errors is not easy, in particular due to the lateral averaging brought by the diffraction on
formula (\ref{phase}). Also the phase field is not always determined up to a region where $\phi=0$, inducing
some error in the inverse Abel transform. Thus, to check the accuracy, a direct comparison  was made
with a fiber-optic probe hydrophone for the pressure at the focus. A fiber-optic probe hydrophone
\cite{staudenraus1993,davitt2010} determines the liquid refractive index modulation, by measuring the
reflection coefficient $R$ at the tip of an optical fiber. Reflection coefficient $R$  is given by the
Fresnel formula:
\begin{equation}\label{fresnel}
R=\left[ \frac{n_f-(n_w + \delta n_w)}{n_f + (n_w + \delta n_w)}\right]^2
\end{equation}
where $n_f$, $n_w$ are the optical fiber and water refractive indices. Water refractive index modulations
$\delta n_w$ due to sound waves are retrieved from those of $R$, after averaging over typically $100$~bursts. 
Corrections due to the non zero compressibility of the fiber core are taken into account by:
\begin{equation}
\delta n_f=\delta n_w \frac{\partial n_f/\partial P}{\partial n_w/\partial P}.
\end{equation}
The derivatives of refractive indices with respect to pressure are $\partial n_w/\partial P \simeq 1.4\times
10^{-4}$~MPa$^{-1}$ \cite{iapws1997} and $\partial n_f/\partial P \simeq 1.5\times10^{-5}$~MPa$^{-1}$ \cite{davittpc}.
The correction amounts to about $7$~\%.

The physical quantity probed by the fiber-optic probe hydrophone and the interferometer being the same, a
direct comparison can be made \cite{note2}. The fiber tip was set $0.12$~mm above the piezo hemisphere center. This
position was determined accurately by the images acquired during the measurement. Refractive index
variations at the same point were also computed from interferometric measurements, taken with the
hydrophone removed. Results are presented in figure 5.
Both measurements are compatible within 5~\%. This is to be compared to the reproductibility of the hydrophone
measurements (typically 10~\%), and the uncertainties in the inverse Abel transform due to the incomplete phase maps
(which can amount to 5~\%). Hence the agreement is satisfactory.

\section{Conclusion}\label{sec:7}
It appears that multiphase interferometric imaging can be easily applied to repetitive fast transient phenomena
for which optical phase is a good observable. It provides quickly extensive data compared to point by point
measurements. We have shown in the case of a sound wave that these data are quantitatively reliable. The
multiphase feature brings a better immunity to optical defects in the images and an improved signal to noise
ratio. Phase unwrapping in time appears as a very simple and robust algorithm. The longer acquisition time
was found acceptable. Hence this method could be more widely used than it has been up to now to study fast
transient phenomena.

\section*{Acknowledgements}
This research has been funded by the ERC under the European Community's FP7 {\em Grant Agreement}
n$^\circ$240113 and by the {\em Agence Nationale de la Recherche}, contract 05-BLAN-0084-02 META.

\newpage
\clearpage
\section*{Captions}
\vspace{1cm}

\begin{figure}
\centering
\includegraphics[width=0.8\textwidth]{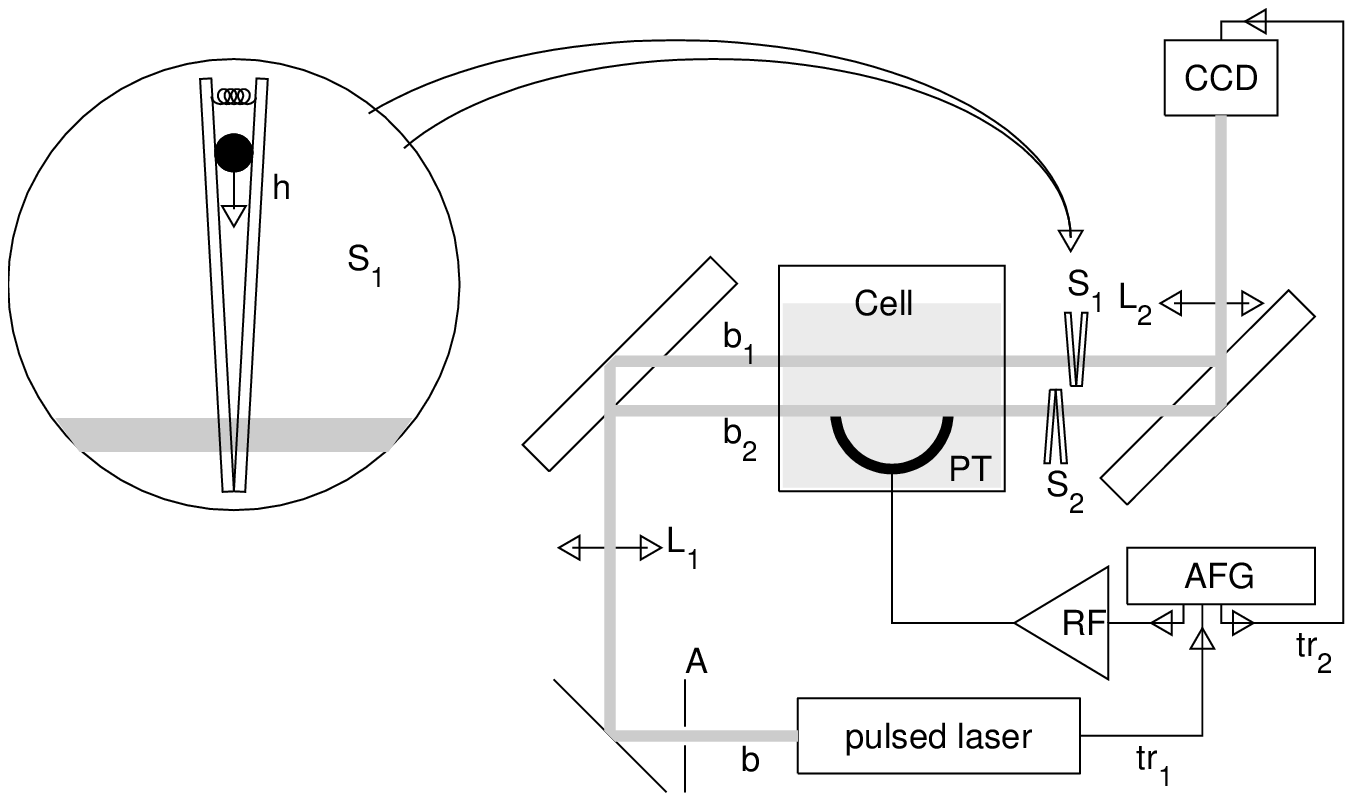}%
\caption{}\label{f1} %
\end{figure}

Figure \ref{f1}: Experimental setup. b: attenuated laser beam, tr1: trigger pulse sent by the laser 170~$\mu$s before the laser pulse, AFG: function generator, tr2: trigger pulse for the CCD camera, PT: piezo transducer driven by the amplifier RF, S$_1$: scanning plates for the interferometer, S$_2$: compensating plates. The inset shows how the optical path for the beam b$_1$ is changed by moving the separator of  S$_1$ plates by $h$.
\vspace{1cm}

\begin{figure}[bt]
\centering
\includegraphics[width=\textwidth]{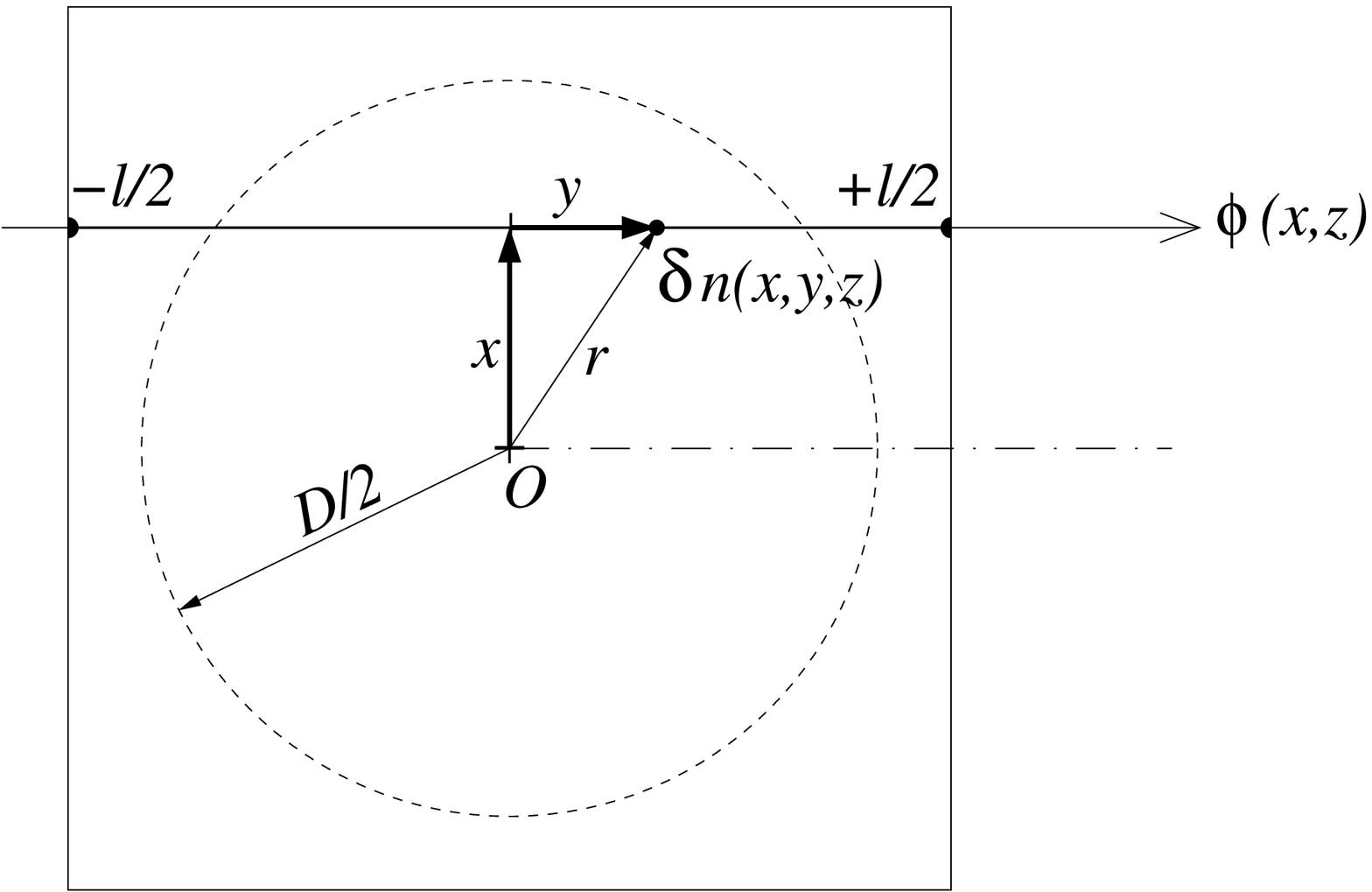}
\caption{}\label{f2}
\end{figure}

Figure \ref{f2}: Measured phase $\phi(x,z)$ results from the integration of the optical phase shift over the %
beam path in the cell. Dashed circle with diameter $D$: limit of the sound field.
\vspace{1cm}

\begin{figure}[tb]
\centering
\includegraphics[angle=-90,width=\textwidth]{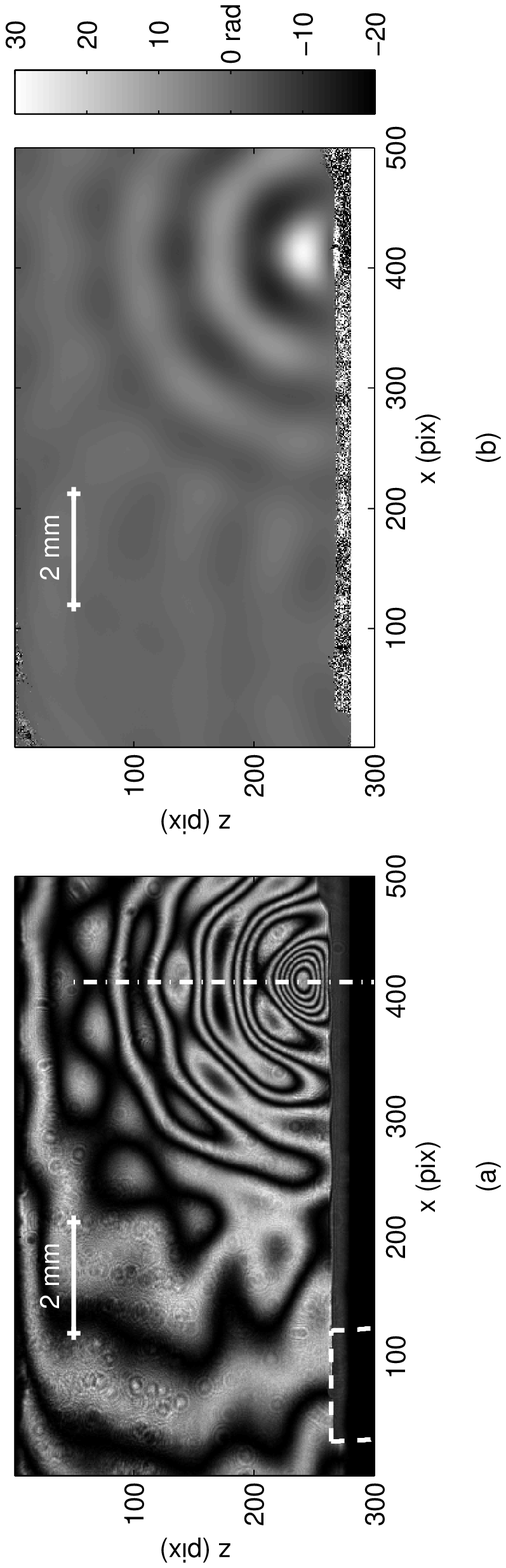}
\caption{}\label{f3}
\end{figure}

Figure \ref{f3} : {\it a)} Image of the interference field above the piezo hemisphere. Dash-dotted white line %
shows its axis and the dashed line outlines the profile of its meridian section. %
{\it b)} Phase field determined from 25 similar images with stepped optical phase. Non transparent %
regions of the field of view appear as random numbers.
\vspace{1cm}

\begin{figure} %
\centering
\includegraphics[angle=-90,width=\textwidth]{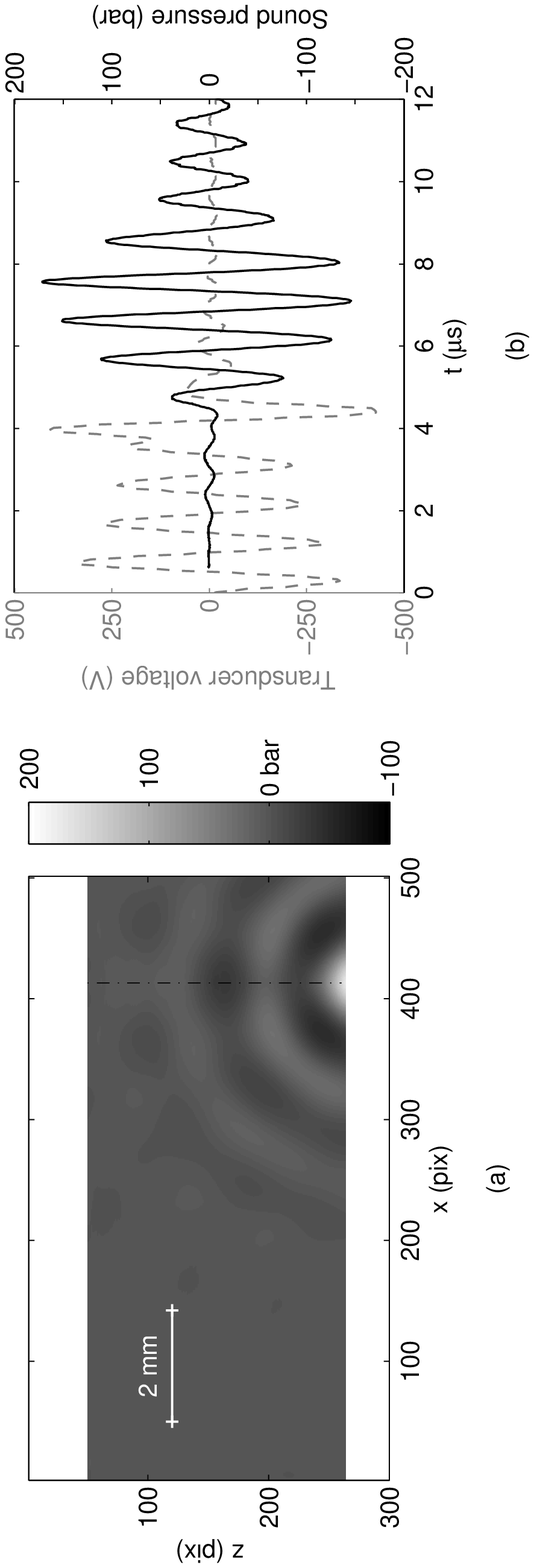}
\caption{}\label{f4}
\end{figure}

Figure \ref{f4} : {\it a)} Pressure map computed from the phase map Fig. 3.{\it b}  by Abel inversion from pixel 1 to 414. Area from $x=414$ to $x=500$ %
are filled by symmetry. {\it b)} Time variations of the computed pressure at the focus while the sound pulse goes through (solid line). %
The transducer excitation voltage is also plotted (dashed line), starting at $t=0$. 
\vspace{1cm}

\begin{figure} %
\centering
\includegraphics[width=0.5\textwidth]{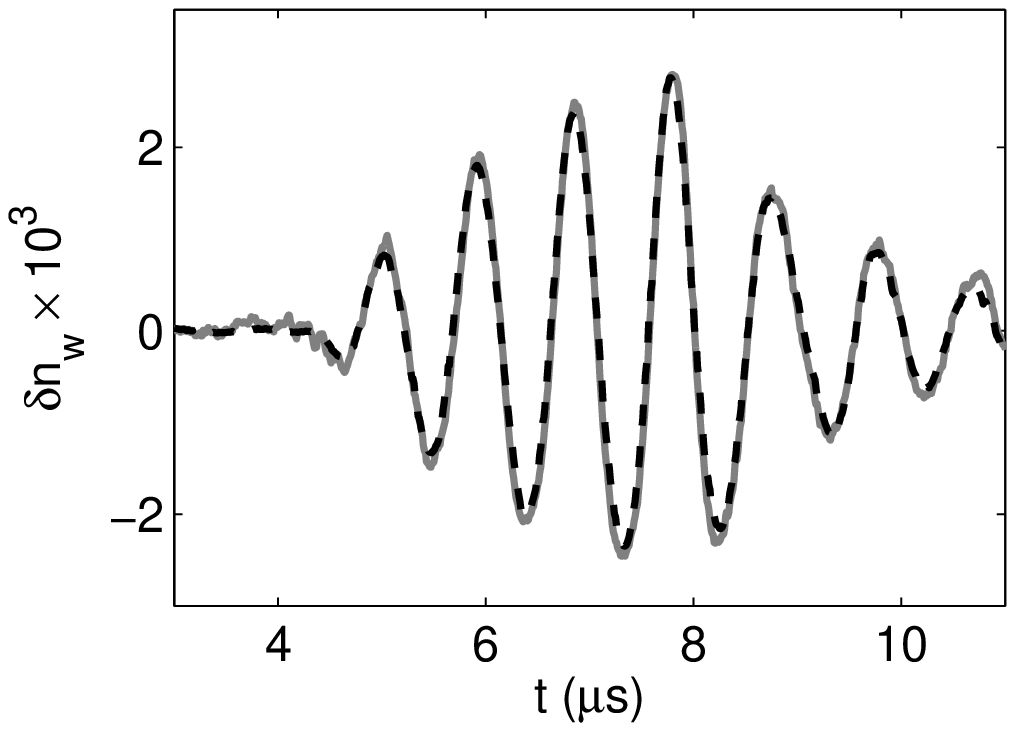}
\caption{}\label{f5}
\end{figure}

Figure \ref{f5} : Comparison of refractive index modulations obtained from the fiber-optic probe hydrophone (gray continuous line) and from inverse Abel transform of the phase map (dark dashed line).

\newpage
\clearpage

\newpage
\clearpage

\newpage
\clearpage

\newpage
\clearpage

\newpage
\clearpage

\begin{thebibliography}{99}
\bibitem{bt} The apparatus is in fact designed for working with low temperature samples protected by four windows on each
side.
\bibitem{note2} The fact that the lasers used in both methods have different wavelengths ($532$~nm and $808$~nm)
introduces a negligible correction.
\end{thebibliography}
\end{document}